\documentclass[aps,prb,twocolumn,a4paper,floatfix,superscriptaddress,showpacs,showkeys]{revtex4-1}

\usepackage[usenames,dvipsnames]{color}

\usepackage{dcolumn,amsmath,xspace}
\usepackage{graphicx,color}

\usepackage{epstopdf}

\newcommand{\degC}{\ensuremath{\,^{\circ}\text{C }}}

\newcommand{\sixrt}{\ensuremath{(6\sqrt{3}\!\times\!6\sqrt{3})\text{R}30^\circ}~}

\begin{document}
\title{Edge states and ballistic transport in zig-zag graphene ribbons: the role of SiC polytypes.}


\author{A.L. Miettinen}
\author{M.S. Nevius} 
\affiliation{The Georgia Institute of Technology, Atlanta, Georgia 30332-0430, USA}
\author{W. Ko}
\author{M. Kolmer}
\author{A.-P Li}
\affiliation{Center for Nanophase Materials Sciences, Oak Ridge National Laboratory, Oak Ridge, Tennessee 37831, USA}
\author{M. N. Nair}
\author{A. Taleb-Ibrahimi}  
\affiliation{Synchrotron SOLEIL, L'Orme des Merisiers, Saint-Aubin, 91192 Gif sur Yvette, France}
\author{B. Kierren}
\author{L. Moreau} 
\affiliation{Institut Jean Lamour, CNRS-UniversitŽ de Lorraine, 54506 Vandoeuvre les Nancy, France}
\author{E.H. Conrad}\email[email: ]{edward.conrad@physics.gatech.edu}
\affiliation{The Georgia Institute of Technology, Atlanta, Georgia 30332-0430, USA}
\author{A. Tejeda}
\affiliation{Laboratoire de Physique des Solides, Universit\'{e} Paris-Sud, CNRS, UMR 8502, F-91405 Orsay Cedex, France}
\affiliation{Synchrotron SOLEIL, L'Orme des Merisiers, Saint-Aubin, 91192 Gif sur Yvette, France}

\begin{abstract}
{
Zig-zag edge graphene ribbons grown on 6H-SiC facets are ballistic conductors.  It has been assumed that zig-zag graphene ribbons grown on 4H-SiC would also be ballistic. However, in this work we show that SiC polytype matters; ballistic graphene ribbons only grow on 6H SiC.  4H and 4H-passivated ribbons are diffusive conductors.  Detailed photoemmision and microscopy studies show that 6H-SiC sidewalls zig-zag ribbons are metallic with a pair of n-doped edge states associated with asymmetric edge terminations, In contrast, 4H-SiC zig-zag ribbons are strongly bonded to the SiC; severely distorting the ribbon's $\pi$-bands.  $\text{H}_2$-passivation of the 4H ribbons returns them to a metallic state but show no evidence of edge states. 
}
\end{abstract}

\maketitle
\newpage

\begin{figure}
\includegraphics[angle=0,width=8cm,clip]{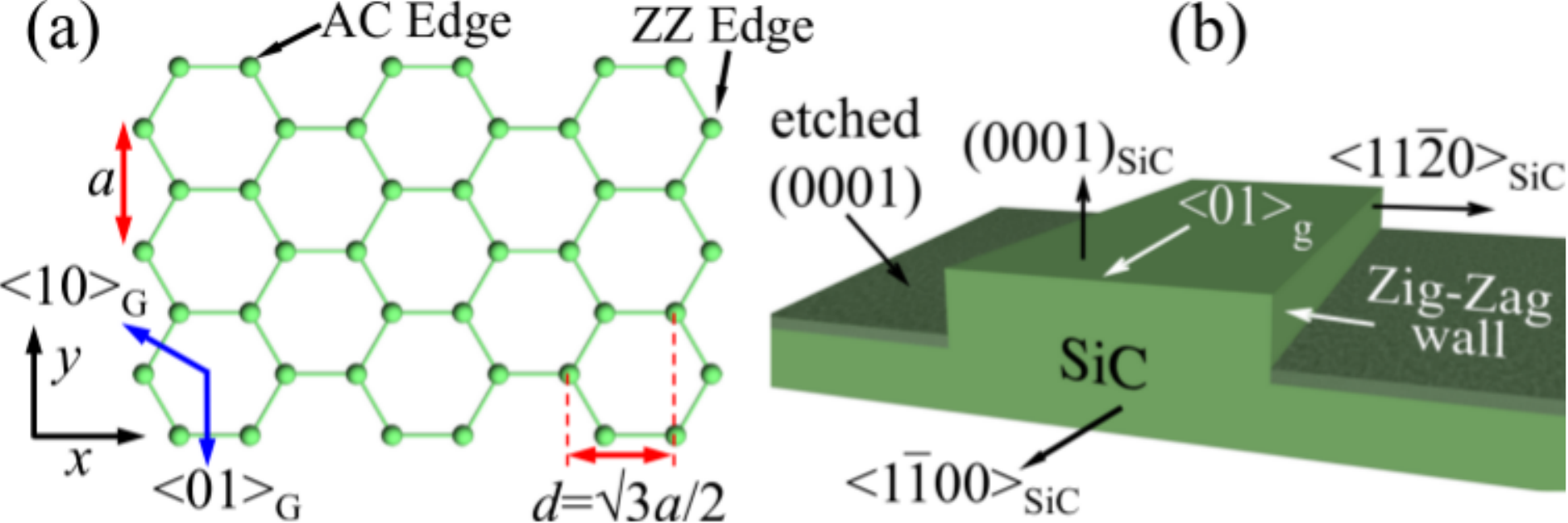}
\caption{(a) Graphene lattice. (b)  The pre-growth etched SiC(0001) step geometry to grow ZZ-edge sidewall graphene. The distance between ZZ rows is $d=\sqrt{3}a/2$, $a=2.462$\AA.
} \label{F:Step_Geom}
\end{figure}

Epitaxial graphene (EG) is graphene grown from silicon carbide (SiC).\cite{Berger_JCP_04}  It has a known orientation relative to the SiC substrate and can be grown as uniform single layers.  The bottom-up growth of EG ribbons on facets of patterned SiC(0001) shallow trenches (known as ``sidewall" graphene) was proposed as a viable route towards graphene electronics;\cite{Nakada_PRB_96,Berger_JCP_04}  circumventing patterned exfoliated graphene's lithographic limits on ribbon width and edge disorder.\cite{Sprinkle_NatNano_2010,Todd_NL_09,Ritter_NatMat_09,Han_PRL_10,Mucciolo_PRB_09,Sols_PRL_07} This is because the edges of EG ribbons are defined entirely by the orientation of the SiC(0001) pre-growth trenches. Trenches parallel to the SiC $\langle 1\bar{1}00\rangle$ direction produce zig-zag (ZZ) edges ribbons on the SiC step facets [see Fig.~\ref{F:Step_Geom}].  Armchair (AC) edge ribbons grow on steps parallel to the SiC $\langle 11\bar{2}0\rangle$ direction. 

An exciting work found that ZZ-edge sidewall ribbons grown on 6H-SiC substrates were room temperature ballistic conductors using 2- and 4-point measurements.\cite{Baringhaus_JPCM_13,Baringhaus_Nature_14} The current development of ballistic devices on modern 4H-substrate has implicitly assumed that ZZ-edge graphene grown on 4H- and 6H-SiC would be the same. 
However, attempts to measure the electronic structure of 4H ZZ-edge graphene, using similar growth methods as in Ref.\,[\onlinecite{Baringhaus_Nature_14}], found no evidence of metallic 4H sidewalls graphene despite exploring growth conditions up to the melting point of the SiC trenches.\cite{Nevius_phd}  These conflicting results lead to the unresolved question: what structural or growth variables affect ZZ-edge sidewall graphene formation?
 In this work, we show that the dominant factor in ZZ-edge sidewall graphene growth is the SiC polytype, not the growth method.  Angle resolved photoemission (ARPES) measurements show that sidewall ZZ-ribbons with metallic $\pi$-bands readily grow on 6H-SiC but not on 4H-SiC.  On 4H-SiC, the graphene's $\pi$-bands are severely distorted by graphene-Si bonds to the SiC facets; similar to the graphene-substrate bonding that makes the first graphene layer on SiC(0001) semiconducting.\cite{Emtsev_PRB_08,Nevius_PRL_15,Nair_NL_17,Conrad_NL_17}  $\text{H}_2$-passivation of 4H-ribbons restores the $\pi$-band's metallic character. The 6H ZZ-edge ribbons have two flat bands 
 below the Fermi Energy ($E_F$).  These bands are consistent with the nearly flat edge states predicted for ZZ-ribbons with asymmetric edge terminations.\cite{Deng_Car_14}  The broken symmetry of the 6H-edge states has the potential to be used in spin valve devices.\cite{Deng_Car_14}   
4H-passivated ribbons show no evidence of edge states.  This is corroborated by the fact that only 6H ribbons are ballistic while both 4H and passivated 4H ribbons are diffusive conductors.  

\begin{figure}
\includegraphics[angle=0,width=8.0cm,clip]{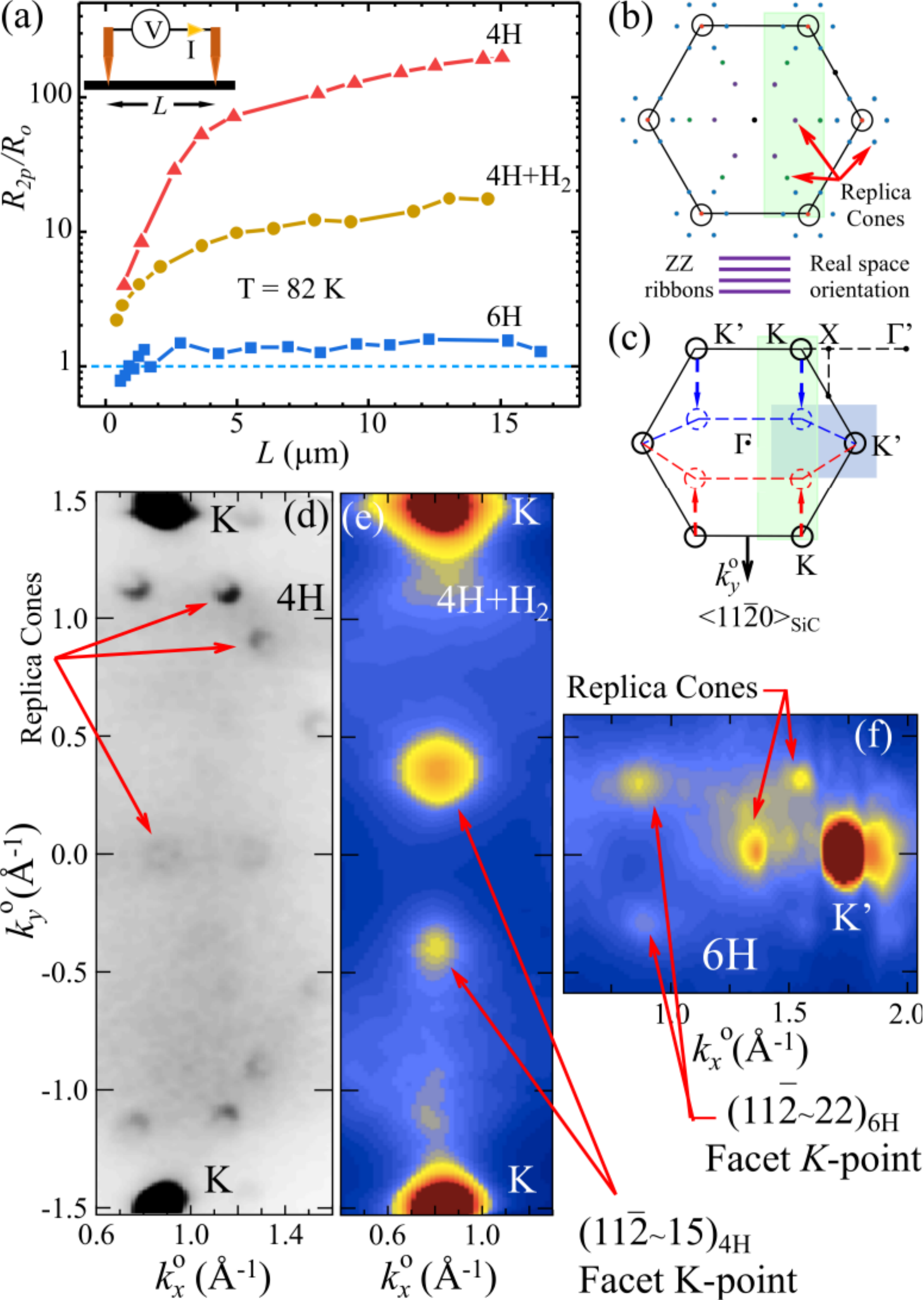}
\caption{
(a) Normalized 2-point resistance vs probe separation for three different ZZ-ribbons: 4H, 4H+H-passivated and 6H ($R_o\!=\!h/e^2$).
(b) and (c) Graphene (0001) BZ. Dots in (b) are $6\!\times\!6$ replica cone positions. 
(c) The compressed graphene BZs (dashed lines and circles) of the $\{11\bar{2}n\}$ (blue) and $\{\bar{1}\bar{1}2n\}$ (red) plotted in the (0001) coordinate frame.  
(d), (e), and (f) are ARPES constant $E$ cuts for three ZZ-sidewall samples ($E\!-\!E_F\!=\!0.09$\,eV and $h\nu\!=\!36$\,eV).  
(d) A cut for 4H ZZ-ribbons [green area in (b) and (c)] showing intensity from both (0001) and $(6\!\times\!6)$ replica cones. 
(e) Same as (d) but after $\text{H}_2$-passivation.  Intensity from $\{11\bar{2}\sim\!\!15\}_\text{4H}$ facets Dirac cones become visible.
(f) A cut for 6H ZZ-ribbons [blue area in (c)].  The $\{11\bar{2}\sim\!\!22\}_\text{6H}$ graphene facets cones  are marked. 
} \label{F:FS_ZZ} 
\end{figure}

Figure \ref{F:FS_ZZ}(a) shows 2-point resistance ($R_\text{2p}$) measurements on different sidewall ZZ-ribbons.  The figure shows that $R_\text{2p}$ for 6H-ribbons is independent of probe separation with a value of $R_\text{2p}\!=\!h/e^2$, i.e., they are ballistic conductors like previously measured 6H ZZ-ribbons.\cite{Aprojanz_NatCom_18}  4H and 4H-passivated ribbons on the other hand show diffusive resistance curves.  Similarly, $dI/dV$ measurements of the different sidewall graphene show that 4H-ribbons are gapped semiconductors [see Supplemental material].  As we now discuss, the reason for these transport differences is the nature of the graphene-substrate interaction on the different SiC polytypes .


Figure \ref{F:FS_ZZ} also compares ARPES intensity cuts, $I(\vec{k}^o,E)$ at fixed binding energy (BE$=\!\!E\!-\!E_F$), for three different ZZ-ribbon arrays: 4H-, 4H-$\text{H}_2$-passivated, and 6H-ribbons.  $\vec{k}^o$ is in the (0001) surface plane. The facet walls are sufficiently well ordered after graphene growth so that the $40\,\mu$m diameter ARPES beam, averaging over $\sim\!\!100$ sidewalls, give a good ensemble average of the ribbon band structure.\cite{Hick_NP_2013}   The ARPES intensity is related to the 2D-band dispersion $E(\vec{k}_\parallel)$ where $\vec{k}_\parallel$ is in the plane of the local surface.  Because of the ARPES beam size, bands from both the (0001) and the opposing $\{11\bar{2}n\}$ and $\{\bar{1}\bar{1}2n\}$ facets are simultaneously measured [see Fig.~\ref{F:FS_ZZ}(c)]. Any Dirac cones from graphene on the facets appear shifted in $k^o_y$ relative to the cones of the (0001) Brillouin zone (BZ) [see Fig.~\ref{F:FS_ZZ}(c)].\cite{Hick_NP_2013}   When we attempt to grow ZZ-ribbons on 4H trenches, the $I(\vec{k}^o_\parallel,E)$ map in Fig.~\ref{F:FS_ZZ}(d) only shows K-point cones and $6^\text{th}$-\,order replica cones associated with the reconstructed graphene-SiC(0001) surface [see Fig.~\ref{F:FS_ZZ}(b)].\cite{Bostwick_NatPhys_07,Nevius_PRL_15}  There is no evidence of rotated Dirac cones from graphene on the facets.  Either no graphene has grown or the graphitic carbon that did grow is bonded strongly enough to the SiC facet to significantly distort the graphene's $\pi$-bands.\cite{Nevius_NL_14} The lack of Dirac cones on 4H-sidewalls persists up to  temperatures where the SiC steps melt.\cite{Nevius_phd}  
It was suggested that the lack of 4H ZZ-ribbon Dirac cones was due to disorder.\cite{Epigraphene} 
 However, as we now show, 4H ZZ-ribbons cones appear once the graphene-SiC bonding is broken. 
   

To demonstrate that graphene is strongly bonded to the 4H sidewalls, we have $\text{H}_2$-passivated the 4H-ribbons in Fig.~\ref{F:FS_ZZ}(d). $\text{H}_2$-passivation is known to break the graphene-substrate silicon bonds; turning a semiconducting graphene film on SiC(0001)\cite{Emtsev_PRB_08,Nevius_PRL_15} to a metallic film.\cite{Riedl_PRL_09}  Figure \ref{F:FS_ZZ}(e) shows the same ARPES map as Fig.~\ref{F:FS_ZZ}(d) but after $\text{H}_2$-passivation. The passivated sample shows that a set of modified Dirac cones appearing along the line between the two K-points of the (0001) surface.  As shown in the schematic BZ in Fig.~\ref{F:FS_ZZ}(c), these cones are from graphene on the tilted facets.  We note that the different facet cone intensities are due to ARPES matrix element effects caused by the different angles between the photon polarization vector and the opposing facet normals.  The angle between the (0001) plane and the facet normal, $\theta_F$, is found from the $\vec{k}^o_y$ positions of the facet cones [see supplementary material]. We find  $\theta_F\!=\!23.6\pm0.5^\circ$, corresponding to $\{11\bar{2}\sim\!\!15\}_\text{4H}$ planes. 

 While graphene-Si  bonding causes 4H ZZ-sidewall graphene to be non-metallic, graphene grown on 6H-SiC $\{11\bar{2}n\}_\text{6H}$ facets is naturally metallic. Figure \ref{F:FS_ZZ}(f) shows a partial ARPES map for ZZ-ribbons grown on 6H-SiC. Unlike 4H-ribbons, Dirac cones from 6H-facets appear after growth without passivation.  The 6H-facets have $\theta_F\!=\!24\pm\!0.5^\circ$, corresponding to graphene ribbons on $\{11\bar{2}\!\sim\!\!22\}_\text{6H}$ planes. 
The fact that the $\pi$-bands are observed without H-passivation demonstrates that 4H ZZ-sidewall graphene is bonded very differently to the substrate compared to 6H-ribbons.  Our results suggest that there are more Si dangling bonds available to interact with the sidewall graphene on 4H- compared to 6H-facets.  It is worth noting that the assumption of 4H- and 6H-SiC $\{11\bar{2}n\}$ planes having similar structures and graphene  bonding is contradicted by earlier SiC growth studies. Experimental and calculated surface free energies of 4H and 6H-SiC 
 ZZ $\{11\bar{2}n\}$ planes clearly show that only 6H-SiC is expected to have a stable ZZ-facet [see supplemental material].\cite{NORDELL_MSE_99} 
   
   \begin{figure}
\includegraphics[angle=0,width=8cm,clip]{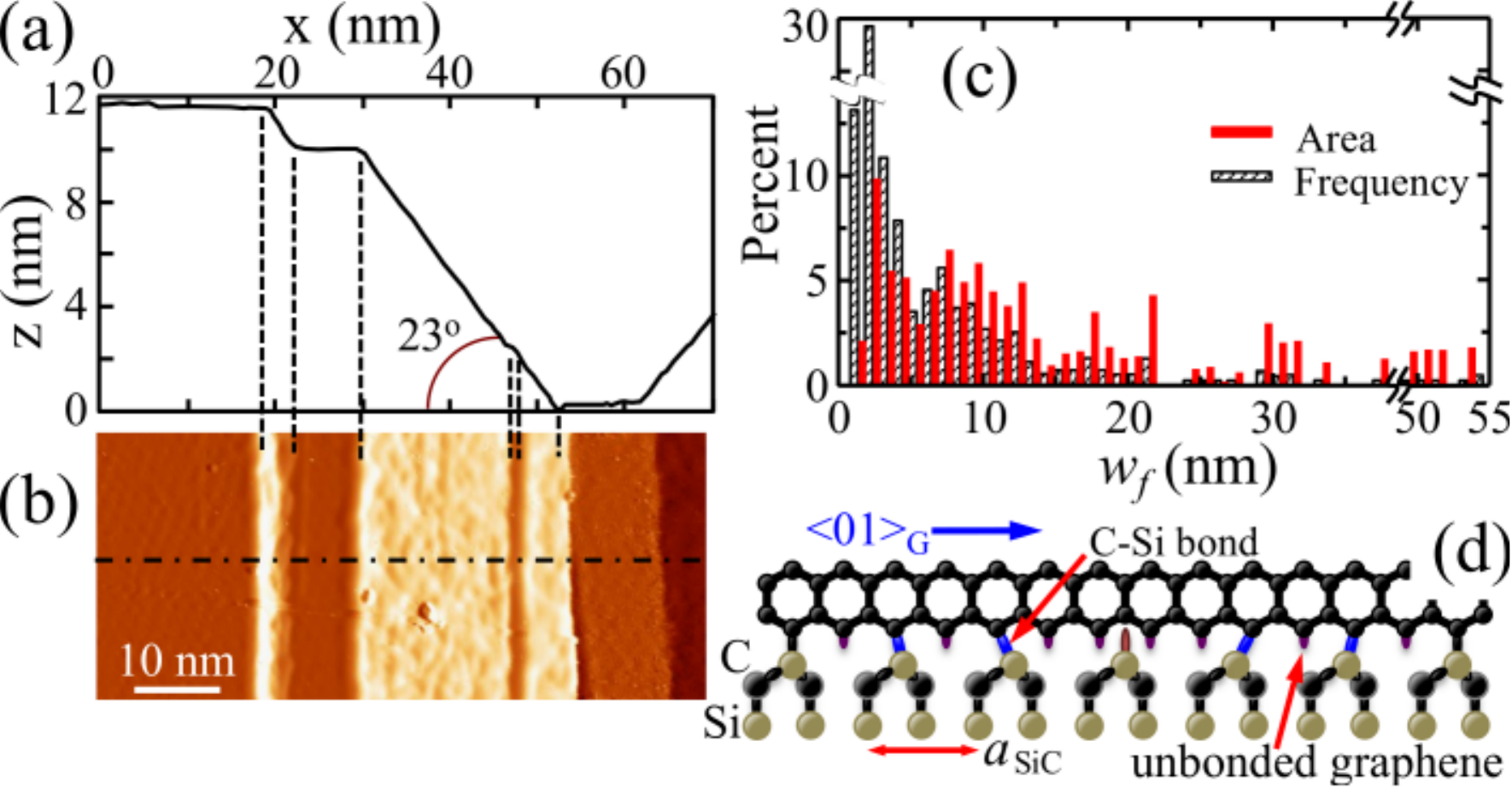}
\caption{(a) Post graphene growth profile near the bottom of a 25\,nm high 6H ZZ-edge steps.  (b) A $dI/dV$ image of the step in (a).  (c) Histogram of 250 facet widths plotted by both relative frequency and  areal coverage. (d) Bonding geometry at a ZZ-edge ribbon to a commensurate bulk terminated \sixrt SiC surface. 
} 
\label{F:STM}
\end{figure}

 Unlike AC-steps where a single $(1\bar{1}07)$ facet covers $\sim\!\!70\%$ of the step area,\cite{Palacio_NLet_15}
6H ZZ-steps have a complicated facet structure.\cite{Baringhaus_JPCM_13,Aprojanz_NatCom_18} The ZZ-steps consist of many $\{11\bar{2}\!\sim\!\!22\}_\text{6H}$-(0001) plane pairs [see Figs.~\ref{F:STM}(a)].  The $\{11\bar{2}\!\sim\!\!22\}_\text{6H}$ facets have a broad width distribution as shown in Fig.~\ref{F:STM}(c). The histogram gives an average 6H facet width of $\bar{w}_f\!\sim\!6\pm8$\,nm with a high number of 1-2\,nm facets.  
The important question is how the graphene ribbon width-distribution, $N(W_r)$, is related to the facet width distribution $N(w_f)$, i.e. does the facet graphene flow onto and over the (0001) nano-terraces as a continuous very wide ribbon (like a carpet draping over a staircase) or does the facet graphene terminate somewhere on an adjacent (0001) terraces so that
$N(W_r)\!\sim\!N(w_f)$? As we will show, both STM and ARPES find that the graphene ribbon width is similar to the facet width. 

 \begin{figure*}[htb]
  \includegraphics[angle=0,width=15.5cm,clip]{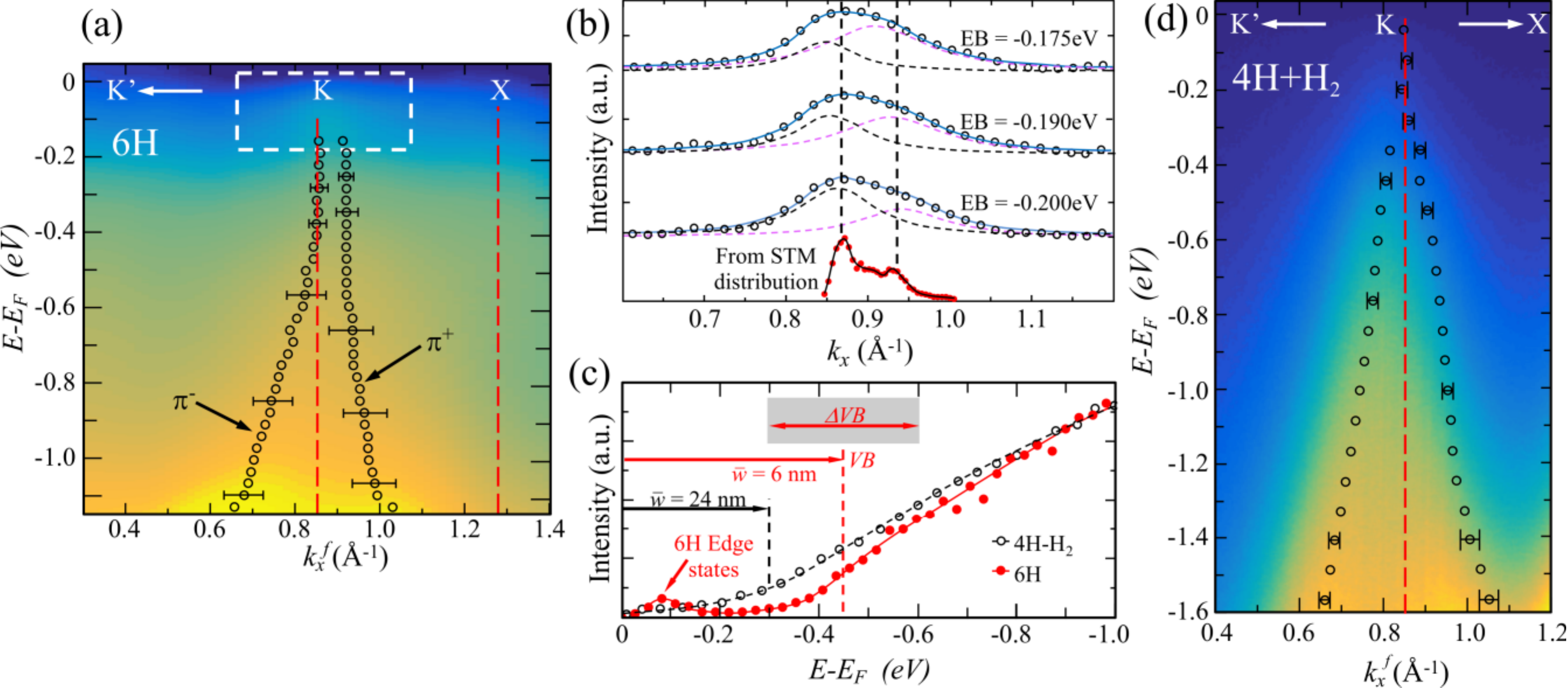}
\caption{
(a) ARPES measures bands of 6H-ZZ sidewall ribbons along the K'K$\Gamma'$ direction (parallel to the ZZ-edge).  $T_\text{sample}=100$\,K.  $k^f_x$ is in the plane of the $(11\bar{2}\sim\!\!22)$ facet. Circles ($\circ$) mark the band  positions. 
(b) Typical MDC fits showing the $\pi$-bands asymmetry for EB$<\!-0.4$\,eV. Solid red circles show the predicted asymmetry from the measured facet $N(w_f)$ distribution and Eq.~(\ref{E:Kpoint}).
(c) Integrated $\pi$-band intensity for 6H (red circles) and $\text{H}_2$-passivated 4H ribbons (black circles).  Red arrows mark the estimated valance band edge and width from 6H STM ribbon $N(w_f)$.
(d) ARPES bands from H$_2$-passivated 4H-ZZ ribbons along the K'K$\Gamma'$ direction (30\,nm steps). $k^f_x$ is in the $(11\bar{2}\sim\!\!15)_\text{4H}$ plane. 
 }
\label{F:ZZ_Cones}
\end{figure*}

  Figure \ref{F:STM}(b) shows a $dI/dV$ map of the step profile in Fig.~\ref{F:STM}(a). The fact that the facets are bright compared to the (0001) nano-terraces indicates that there is a discontinuity in the electronic structure of the graphene on facet and the semiconducting graphene that is known to grow on the nano-terraces.\cite{Palacio_NLet_15}  These results suggest that the facet graphene either terminates into the SiC(0001) surface or transitions into a semiconducting form of graphene on the (0001) nano surface.  In either case, the results suggest that graphene is a metallic ribbon of width proportional to the facet width.  As we will show below, this statement is supported by the ARPES data.  


Both the 6H- and 4H-passivated ZZ ribbons are sufficiently ordered to allow detailed area-averaged band  measurement. Figure \ref{F:ZZ_Cones}(a) shows the 6H ribbons' band intensity for $k^f_x$ along the K'K$\Gamma'$ direction of the $(11\bar{2}\!\sim\!\!22)$ facet plane. The 6H ZZ-ribbon's $\pi^-$ and $\pi^+$ bands' momentum and widths were determined from Lorentzian fits to momentum distribution curves (MDC); $I(k^f_x,BE)$ at fixed BE [see Supplemental material]. The $\pi$-band positions (marked by circles) show a distorted Dirac cone.  For BE$>\!-0.4$\,eV, the bands have an asymmetric dispersion with a larger band velocity, $v_F$, for the $\pi^+$ band ($v_F^+/v_F^-\!\sim\!1.7$). Both tight binding (TB) and ab initio models predict this asymmetry for narrow ribbons. \cite{Yang_PRL_07,Wakabayashi_STAM_10}  

For BE$<\!-0.4$\,eV, the MDC fits show a consistent apparent spitting of $\pi$-bands [see Figs.~\ref{F:ZZ_Cones}(a) and (b)].  
While the splitting appears unusual, it is a result of a distribution of the valance band maximum ($\text{VB}_\text{m}$) positions ($k^m_x$) from a ribbon width distribution on the facets that contains a large number of sub 5\,nm parallel ribbons.  
To demonstrate this, we use a TB model for the ribbon's band structure. In this model the $n\!=\!0$ subband is due to the ribbon edges (the edge state).  The $k^f_y$ wavevector for this state is imaginary, localizing the wavefunction to the edges for $k_c\leq k^f_x\leq \text{X}$, where the critical momentum $k_c$  depends on ribbon width, $W_r$:\cite{Wakabayashi_STAM_10}    
 \begin{equation}
  k_c=\frac{2}{a}\arccos{\lbrack\frac{1}{2}\frac{W}{W+d}\rbrack
 }
\label{E:Kpoint}
\end{equation}
$d$ is the spacing between ZZ chains [see Fig.~\ref{F:FS_ZZ}]. In both TB and first principle models, the top of the $n\!=\!1$ subband corresponds to the ribbon's $\text{VB}_\text{m}$.\cite{Yang_PRL_07}  To a very good approximation, $\text{VB}_\text{m}$ occurs at $k_x^\text{m}\!\sim\!k_c$ [see Supplemental material]. For ribbons with $W\!\gg\!d$, $\text{VB}_\text{m}$ occurs at the K-point.  Ribbons with $W\!\sim\!d$ have $\text{VB}_\text{m}$ shifter to higher $k_x$.  Figure \ref{F:ZZ_Cones}(b) compares the calculated $k_c$ position from Eq.~(\ref{E:Kpoint}) using the experimental $N(w_f)$ distribution in Fig.~\ref{F:STM}(c).  We have convoluted the calculated $k_x$ with a $\Delta k_x\!=\!0.05\,\text{\AA}^{-1}$ window consistent with the measured Lorentzian width. The calculated $\text{VB}_\text{m}$-distribution reproduces the asymmetric ARPES intensity very well. This can only happen if $N(W_r)\!\sim\!N(w_f)$, i.e. $W_r\sim\!w_f$. 
 
The equality of the facet and graphene ribbon widths also explains the VB's intensity decay and the $\pi$-bands' momentum broadening for BE$>\!-0.4$\,eV,  To show this, we use the calculated $n\!=\!0$ subband energy splitting, $\Delta^\text{o}(W_r)$, at the K-point in the GW approximation;\cite{Yang_PRL_07}  
 \begin{equation}
  \Delta^\text{o}\!\approx \!A/(W_r+\delta), 
\label{E:Delta0}
\end{equation}
where $A\!=\!38$\,eV\AA~and $\delta\!=\!16$\,\AA.\cite{Yang_PRL_07}   Roughly, $\text{VB}_\text{m}$ is $\sim\!0.5\Delta^\text{o}$ below $E_F$ for neutral ribbons. Using the STM measured $w_f$-distribution for $N(W_r)$ in Eq.~(\ref{E:Delta0}) gives $\text{VB}_\text{m}\!=\!0.44$\,eV with $\Delta\text{VB}=0.24$\,eV. These values are marked on the plot of the 6H-ZZ ribbon $\pi$-band intensity, I(BE), in Fig.~\ref{F:ZZ_Cones}(c).  They are in good agreement with the experimental intensity that has a broad decay with an inflection point near 0.5\,eV.

Finally, the $\pi$-band $\Delta k^f_x$ broadening near the inflection point of the integrated 6H $\pi$-band intensity [BE\,$\sim\!-0.45$\,eV in Fig.~\ref{F:ZZ_Cones}(c)] is $\sim\!0.17\,\text{\AA}^{-1}$; three times the broadening at lower BE. The larger $\Delta k^f_x$ near $\text{VB}_\text{m}$ is caused by overlapping sub-band energies from ribbons with different widths.  Again, area-averaged ARPES contains intensity from a distribution of sub-bands, $n(W_r)$ (shifted up and down in BE) that leads to an apparent $\Delta k_x$-broadening of the $\pi$-bands. To estimate $\Delta k_x$, we assume a linear $\pi$-band dispersion, $E\!=\!\hbar\tilde{c}k$, where $\tilde{c}$ is the average measured band velocity of the $\pi$-bands.  If the apparent band broadening is $\Delta E\approx\! \delta\Delta^\text{o}(W_r)$, then  $\Delta  k^f_x$ is given by:
 \begin{equation}
  \Delta  k^f_x\approx\frac{\Delta^\text{o}(\bar{w}_f)\Delta\bar{w}_f}{\hbar\tilde{c}(\bar{w}_f+\delta)},
\label{E:dk}
\end{equation}
where we have again assumed that $w_f\!=\!W_r$.  Using the measured STM values for $\bar{w}_f$, gives $\Delta  k^f_x=0.14\text{\AA}^{-1}$; in good agreement with the measure value.  In short, the $\pi$-band's shape, splitting, and $\Delta  k^f_x$ broadening are all consistent with 4H ZZ-ribbons having a width approximately equal to the $(11\bar{2}\!\sim\!\!22)$ facet widths.

\begin{figure}
\includegraphics[angle=0,width=8cm,clip]{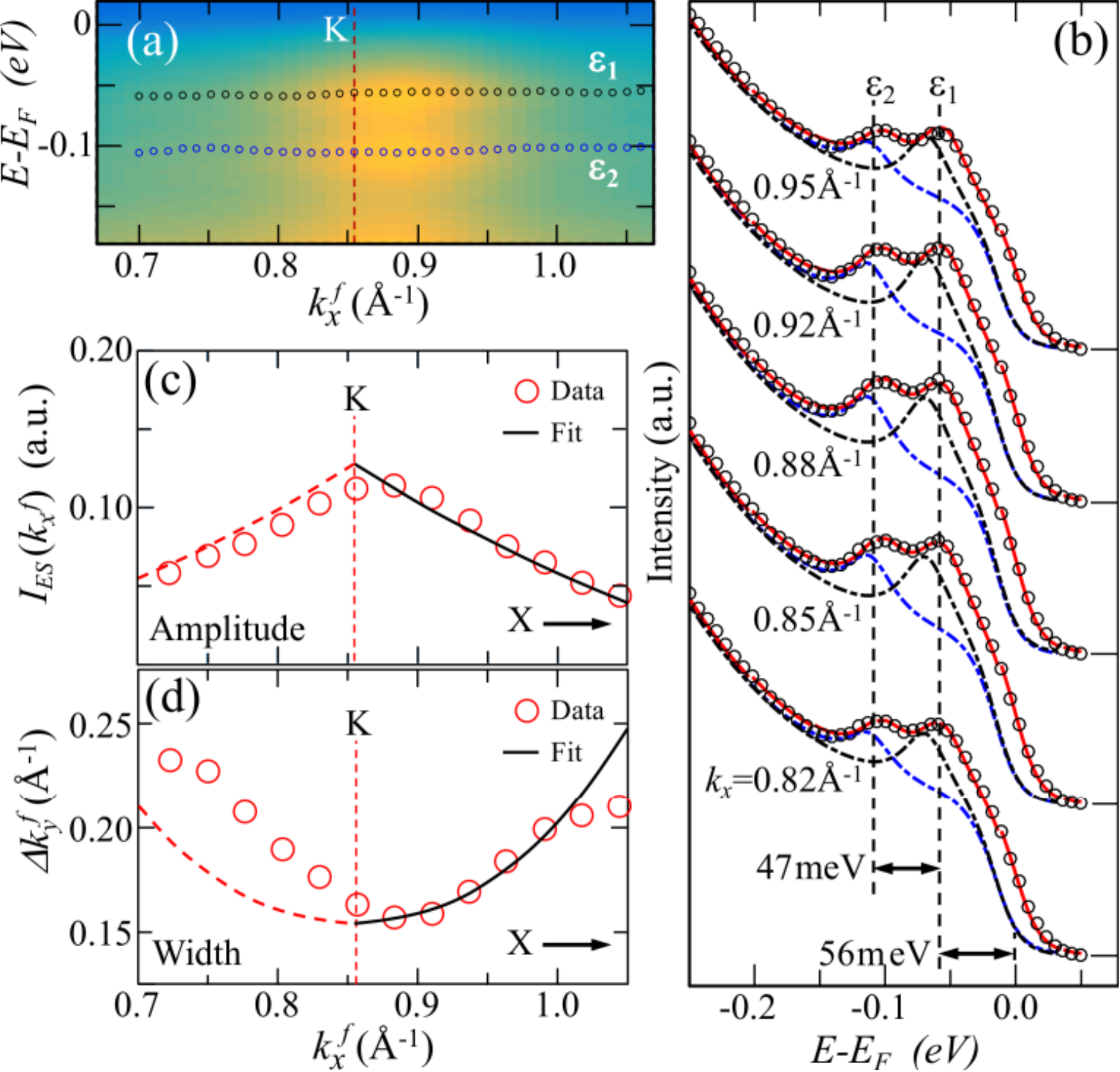}
\caption{
(a) Close up of 6H-ZZ sidewall ribbon band structure near $E_F$  [dashed box in Fig.~\ref{F:ZZ_Cones}(a)] showing flat bands, $\epsilon_1$ and $\epsilon_2$.  Circles $\circ$ mark the peak positions from EDCs fits. 
(b) Sample EDCs near $E_F$ in (a) ($\circ$ are data). 
A two Lorentzian (dashed black and blue lines) plus background and Fermi-Dirac cut off fit is shown (red
Solid line). 
(c) and (d) are $\epsilon_1$ and $\epsilon_2$ average intensity and $\Delta k^f_y$ width, respectively, vs $k^f_x$.
Solid and dashed lines are fits described in text.
  }
\label{F:ZZ_SS}
\end{figure}

 What distinguishes ZZ-ribbons from AC-ribbons is the existence of a ZZ-edge state.\cite{Wakabayashi_STAM_10} Because these states are localized near the ribbon edges, their spectral weight is low.  Nevertheless, we find two states, $\epsilon_1\!$ and $\epsilon_2\!$, near $E_F$ in Fig.~\ref{F:ZZ_Cones}(a) that we can identify as edge states.  The states are seen more clearly in Fig.~\ref{F:ZZ_SS}(a) and (b). Energy distribution curve, EDC, $I(BE\!:\!k^f_x)$ at fixed $k^f_x$, show that the states are essentially dispersionless along KK'$\Gamma'$ [see Fig.~\ref{F:ZZ_SS}(a)].   EDCs near K in Fig.~\ref{F:ZZ_SS}(b) show that the BE of the two states are: $\epsilon_1\!=\!-56$ and $\epsilon_2\!=\!-103$\,meV.
 Their  energy width is 58~meV; essentially the expected thermal broadening for the $T\!=\!100$\,K sample.  We identify these bands as ZZ edge states associated with asymmetric edge terminations.  This assignment follows from three observations:  (i) Their intensity and perpendicular broadening along $\text{K}\!<\!k^f_y\!<\!\text{X}$ is consistent with predictions, (ii) the states do not disperse in either $k^f_x$ or $k^f_y$, and (iii) the bands are narrow in E.

 Figures \ref{F:ZZ_SS}(c) and (d) show that the average  $\epsilon_1$ and $\epsilon_2$ intensity decreases for  $k^f_x>\text{K}$ while their perpendicular band width, $\Delta k^f_y$, increases $k^f_x>\text{K}$. These are the expected trends for edge states in a TB model for ZZ-ribbons.  Using symmetric ZZ edges, the TB edge state's charge density, $\rho(y)$, is completely localized at the edge when $k^f_x\!=\!\pi/a$ (the 1D X-point). At lower $k^f_x$, it become more delocalized until at the K-point  ($k^f_x\!=\!2\pi/3a$) $\rho(y)$ is uniform perpendicular to the edge.\cite{Wakabayashi_STAM_10}  Therefore, the edge state band width, $\Delta k^f_y\!\sim\!2\pi/\Delta y$, is largest at X and smallest at K. Furthermore the edge state intensity, $I_\text{ES}(k^f_x)$, is proportional to $\cos(k^f_xa/2)$.\cite{Wakabayashi_STAM_10}  Thus $I_\text{ES}(k^f_x)$ is a maximum at K and decreases as $k^f_x$ approaches the X-point. 
The TB estimates [see Supplemental material] for $I_\text{ES}(k^f_x)$ and  $\Delta k^f_y$, are plotted in Figs.~\ref{F:ZZ_SS}(c) and (d). Note that $\Delta k^f_y$ has been convoluted with a 0.16\,\AA\, window to match the experimental minimum at K. Dashed lines in Figs.~\ref{F:ZZ_SS}(c) and (d) are mirrored versions of the solids lines about K. 

While there are similarities between the 6H ribbon edge states and TB predictions, there are significant differences. First, the measured edge states are narrow in energy ($\Delta E\!=\!58$\,meV).  Since the edge state splitting, $\Delta^\text{o}$, from symmetrically terminated ribbons depends on $W_r$, we would expect $\epsilon_1$ and $\epsilon_2$ to have a significant $\Delta E$ due to $N(W_r)$. From Eq.~\ref{E:Delta0}, the STM $w$-distribution would give $\Delta E\!\sim\!0.5$\,eV; 9 times larger than measured. Furthermore, the $\epsilon_1$ and $\epsilon_2$ bands are flat while theoretical models for symmetric edge terminations predict a strong dispersion along the K'KX direction, regardless of whether or not they are anti-ferromagnetic (AF) or ferromagnetically (F) coupled.\cite{Son_PRL_06,Yang_PRL_07,Pisani_PRB_07,Jung_PRB_09,Yazyev_RPP_10}  Asymmetric  terminations models, on the other hand, give rise to nearly flat bands near $E_F$.\cite{Lee_PRB_09,Deng_Car_14}  In particular $\text{sp}^2$ termination on one edge and $\text{sp}^3$ on the other, gives rise to nearly flat bands through the entire 1D BZ whose energies are essentially independent of the ribbon width.\cite{Deng_Car_14}  In other words, our edge states are not from symmetric ribbons.

\begin{figure}
\includegraphics[angle=0,width=7.0cm,clip]{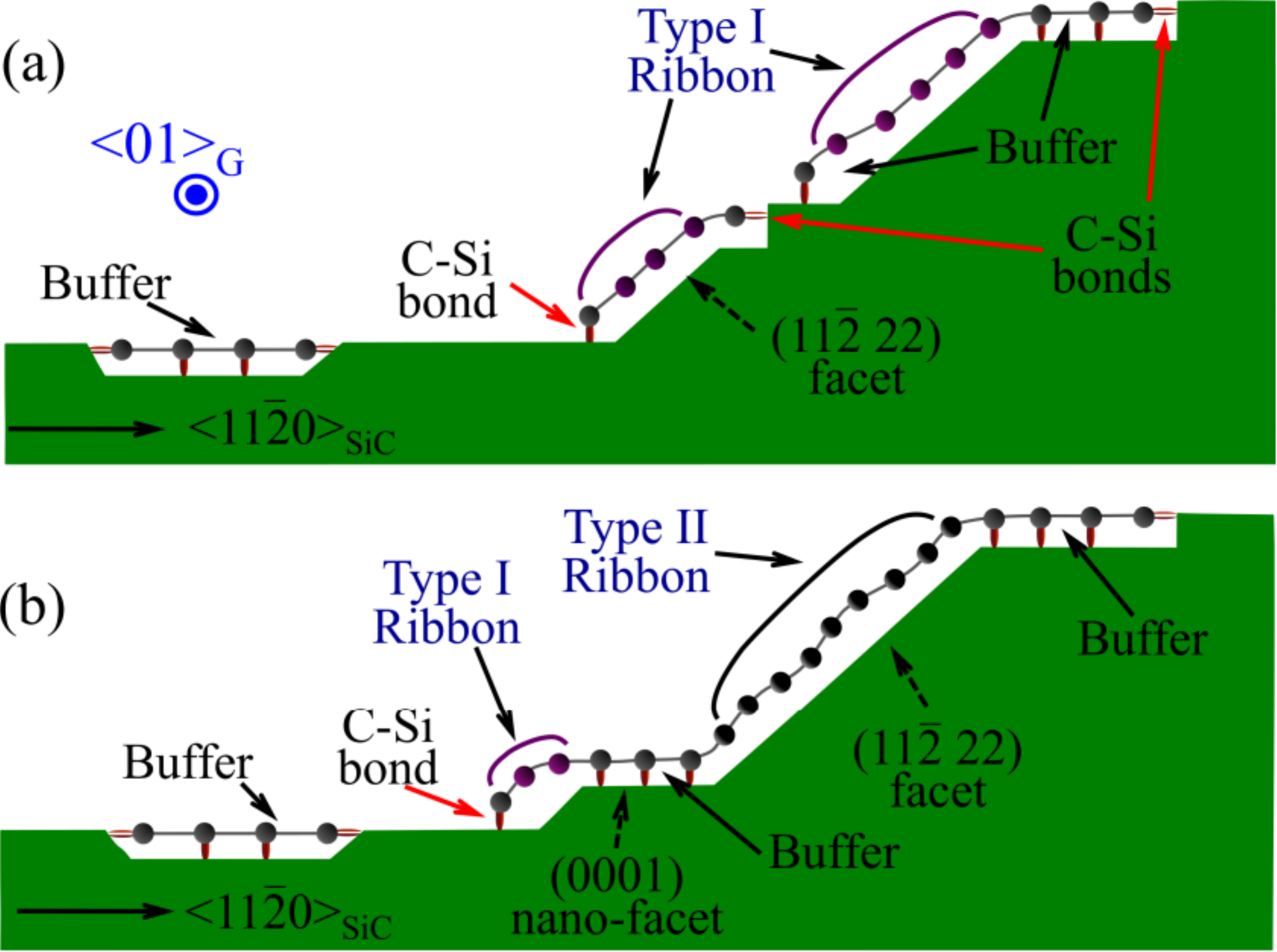}
\caption{Two edge terminated 6H ZZ-ribbons. Functionalized (buffer) graphene is represented by grey circles bonded to the SiC. 
 (a)  Type I ribbons terminated by C-C bonds into buffer graphene on one side and C-Si bonds to the substrate on the other side.  
 (b) A Type II termination with metallic graphene terminated on both sides by C-C bonds into buffer graphene.
 }
\label{F:Termination}
\end{figure}

The fact that the ARPES data points to asymmetric edges is not unexpected.  Based on HRSTEM images of 4H-SiC AC-steps, the ribbons terminate into semiconducting buffer graphene on the macroscopic (0001) surface through $\text{sp}^2$ C-C bonds.\cite{Palacio_NLet_15} At the step bottom, the ribbon terminates by either C-Si $\text{sp}^3$ bonds to the substrate SiC (Type I termination in Fig.~\ref{F:Termination} (a)) or by an intermediate $\text{sp}^2$ C-C bond to buffer graphene on (0001) nano-facets (Type II in Fig.~\ref{F:Termination} (b)). The asymmetric Type I termination is more complicated than Fig.~\ref{F:Termination} indicates. While the ribbon-buffer edge is commensurate and ordered, the C-Si $\text{sp}^3$ edge is incommensurate with the SiC [see Fig.~\ref{F:STM}(g)].\cite{Conrad_NL_17}  The aperiodic C-Si $\text{sp}^3$ bonding leads to $>\!60\%$ bond defects with the edge-carbon either unbonded or re-hybridized in some complicated way. This fraction can be larger since the actual (0001) surface has 20\% less Si than a bulk terminated surface.\cite{Emery_SW_13,Conrad_PRB_17} The edge structure is also complicated by the patterned step edges being slightly rotated with respect to the SiC, $\phi\!\sim2\!-\!2.5^\circ$. This leads to a chirality in the step edges as the graphene terminate into the (0001) planes [see Fig.~\ref{F:STM}(f)].  
Line defects in graphene are also known to lead to flatter band over the entire zone compared to H-terminated ribbons.\cite{Dutta_SREP_15} Regardless of the details of the asymmetric C-Si $\text{sp}^3$ edges, the narrow energy widths and dispersionless character of the observed $\epsilon_1$ and $\epsilon_2$ bands are consistent with edge states from asymmetric edge terminations in the sidewall SiC system. 

 While Type II ribbons resemble a wave-guide geometry with metallic graphene ribbon bonded to a semiconducting graphene on both edges, their terminations are also asymmetric.   This is because buffer graphene on macroscopic (0001) and nano (0001) terraces are electronically different. The dI/dV data in Fig.~\ref{F:STM}(b) clearly show a bias dependent contrast difference between macroscopic and nano (0001) surfaces. Furthermore, only the macroscopic (0001) surface shows the typical $(6\times\!6)$ reconstruction [see Fig.~\ref{F:STM}(e)].\cite{Riedl_PRB_07} How this waveguide affects transport is an open question.  Regardless, Type II ribbon can be thought of as an asymmetric Type I ribbon with a more complicated structure between $\text{sp}^2$ and $\text{sp}^3$ edges.


Data on 4H $\text{H}_2$-passivated ribbons implies a very different ribbon geometry compared to 6H-ribbons.  Figure \ref{F:ZZ_Cones}(d) shows the band structure from the passivated 4H ZZ-ribbons.  The cut through the BZ is the same cut as in Fig.~\ref{F:ZZ_Cones}(a). The flat edge state bands below $E_F$ are not seen in the passivated 4H sample. We would expect a change in the edge states because the hydrogen treatment would not only transform semiconducting buffer graphene  to metallic on all (0001) terraces,\cite{Sforzini_PRL_15} it would also break any C-Si bonds at the Type I edges and hydrogenate a large fraction of unbonded carbon along the edges. While $\text{H}_2$-passivation can p-dope graphene by $\sim\!100$\,meV and shift the states above $E_F$ where they would not be visible in ARPES,\cite{Riedl_PRL_09} transport measurements suggest that no edge states exist.

 
Besides the missing edge states, there are other differences in the 4H-passivated $\pi$-bands compared to 6H-ribbons.  
Unlike 6H-ribbons, the $\text{H}_2$-passivated 4H-ribbon's $\pi^-$ and $\pi^+$-bands are nearly symmetric about the $K$-point with the same band velocity found in macroscopic sheets, $v_F\!\sim\!1\!\times\!10^6$\,m/sec. The $k^f_x$-position of $\text{VB}_\text{m}$ is essentially at the K-point and there is no evidence of a $\Delta k^f_x$ splitting of the $\pi$-bands. This rules out any significant number of sub 5\,nm ribbons. If a $\text{VB}_\text{m}$ exists, it must occur at $\text{BE}\!\lesssim\!-0.3$\,eV.  According to Eq.~\ref{E:Delta0}, that BE would imply a 4H-ribbons width of $\bar{W}_\text{4H}\!>\!24$\,nm. 
It is clear that either the 4H-step is approximately a single $(11\bar{2}\sim\!15)$ facet or the sidewall graphene that grows is only terminated at the top and bottom of the step.
 
 
To summarize, we demonstrate that metallic ZZ-edge epitaxial graphene ribbons only grow on the $\{11\bar{2},22\}_\text{6H}$ facets of the 6H-SiC polytype. While graphene does grow on the 4H polytype, it is bonded to the facet walls in a way that heavily modifies the graphene $\pi$-bands, similar to why the first graphene layer on SiC(0001) is semiconducting because of graphene-SiC bonds.  The non-metallic 4H-ribbons can be turned metallic by H$_2$-passivation that breaks the graphene-sidewall bonds.  

STM, STS, and ARPES measurements show that 6H facet walls consist of multiple $\{11\bar{2},22\}_\text{6H}$-(0001) nano-plane pairs. These pairs appear electronically isolated from each other and give rise to a width distribution where more than $>\!80$\%  of the ribbons are less than 12\,nm wide ($>\!50$\% between 1-3\,nm).  ARPES measurements find two non-dispersing states 56 and 103\,meV below $E_F$ that are identified as ZZ-ribbon edge states.  These states' dispersion, width, and intensity dependence on in-plane momentum indicate that they originate from asymmetrically terminated ZZ-edges.  The lack of an observed crossing of the two states suggests that they are anti-ferromagnetically aligned on opposite edges of the ribbon. Transport measurements shows that these 6H ZZ-ribbons are ballistic conductors up to at least $16\,\mu$m. 


Unlike 6H-ribbons, ARPES shows that the 4H-ribbon appear to be a single wide sheet over the entire 4H facet.  The passivated 4H ribbons show no evidence of n-doped surface states and their transport is diffusive.    
While our results are consistent with the ballistic conduction measured in previous experiments on 6H-SiC sidewall ribbons,\cite{Baringhaus_Nature_14,Aprojanz_NatCom_18}  they contradict the results of non-local transport studies on gated 4H-ZZ sidewall ribbons used to explain the 6H ballistic transport in Ref.~[\onlinecite{Baringhaus_Nature_14}].  We definitively show that 4H ZZ-ribbons are diffusive conductors, not ballistic.  

\vspace*{2ex}
\noindent{\color{red}\textbf{Acknowledgments}}
This research was supported by the National Science Foundation under Grant No. DMR-1401193 (EHC). ALM acknowledges a travel grant from the School of Physics at Georgia Tech.  A.T. acknowledges support from the Agence Nationale de la Recherche (France) under contract CoRiGraph. A portion of the work (multiprobe STM) was conducted at the Center for Nanophase Materials Sciences (CNMS), which is a DOE Office of Science User Facility.

\vspace*{2ex}

\noindent{\color{red}\bf{Methods}}   Samples were prepared starting from a polished SiC(0001) sample from Cree Inc. Trenches with $\{11\bar{2}0\}$ facet walls are formed by e-beam patterning lines on SiC followed by plasma etching to produce well defined 25-30\,nm deep trenches 400\,nm apart over a 1\,mm$^2$ area. The (0001) trench tops are 200\,nm wide. Details of the process are found in the  supplementary material and in Ref.~[\onlinecite{Nevius_phd}]. The ribbons are grown in a confinement-controlled sublimation furnace.\cite{WaltPNAS}  The process causes the surface of the step to reorganize into a complicated set of equilibrium facets and simultaneously grow sidewall graphene. Graphene does not grow well on the SiC(0001) plasma-etched trench bottoms.\cite{Nevius_NL_14} This limits graphene growth to the step walls and a small strip on the (0001) surface at the step edge. To $\text{H}_2$-passivate the post-graphene growth 4H-SiC trenches, samples were heated at 900\degC for 1\,hour in an $\text{H}_2$ furnace ($\text{P}_\text{H2}\!\sim\!800$\,mtorr).  ARPES measurements were done on the high resolution Cassiop\'{e}e beamline.  The line has a total measured instrument resolution of $\Delta E\!\!<\!\!12$\,meV using a Scienta R4000 detector with a $\pm15^{\circ}$ acceptance  at $\hbar \omega\!=\!36$\,eV.  Samples were mounted on a 3-axis cryogenic manipulator.The STM experiments were carried out in an ultra-high vacuum setup with a base pressure in the low $10^{-10}$\,mbar range using a commercial low-temperature Omicron microscope that was modified to minimize capacitive coupling [see Supplemental material and Ref.~[\onlinecite{Didiot_PRB_07}]. Scanning tunneling spectroscopy (STS) and two-probe transport measurements were made on two different cryogenic four-probe scanning tunneling microscope (4P-STM) systems were utilized, one operated at 82\,K and the other at 4.6\,K [1, 2]. 
\cite{Kim_RSI_07,Li_Adv_Func_Mat_13}  All measurements were done under UHV condition ($<\!10^{-10}$\,torr).  Because the graphene sidewall samples were exposed to the air after growth, they were cleaned prior to measurement by annealing in the UHV chamber at 300-500 \degC for a several hours before STM measurements.  See Supplemental material for more details.


\bibliography{refs_ZZ}

\end{document}